\documentclass[prl,floatfix,twocolumn,superscriptaddress]{revtex4-1}
\usepackage{times,graphicx,bbm,amsmath,amssymb,units}
\usepackage{epsfig,color}
\usepackage{hyperref}

\usepackage{pifont}

\newcommand{\ket}[1]{\vert#1\rangle}

\begin{document}
\title{Direct observation of sub-binomial light}
\author{Tim J. Bartley}
\affiliation{Clarendon Laboratory, University of Oxford, Parks Road, OX1 3PU, Oxford, UK}
\author{Gaia Donati}
\affiliation{Clarendon Laboratory, University of Oxford, Parks Road, OX1 3PU, Oxford, UK}
\author{Xian-Min Jin}
\affiliation{Clarendon Laboratory, University of Oxford, Parks Road, OX1 3PU, Oxford, UK}
\affiliation{Department of Physics, Shanghai Jiao Tong University,  Shanghai 200240, PR China},
\author{Animesh Datta}
\affiliation{Clarendon Laboratory, University of Oxford, Parks Road, OX1 3PU, Oxford, UK}
\author{Marco Barbieri}
\affiliation{Clarendon Laboratory, University of Oxford, Parks Road, OX1 3PU, Oxford, UK}
\author{Ian A. Walmsley}
\affiliation{Clarendon Laboratory, University of Oxford, Parks Road, OX1 3PU, Oxford, UK}
\begin{abstract}
Nonclassical states of light are necessary resources for quantum technologies such as cryptography, computation and the definition of metrological standards. Observing signatures of nonclassicality generally requires inferring either the photon number distribution or a quasi-probability distribution indirectly from a set of measurements. Here, we report an experiment in which the nonclassical character of families of quantum states is assessed by direct inspection of the outcomes from a multiplexed photon counter. This scheme does not register the actual photon number distribution; the statistics of the detector clicks alone serve as a witness of nonclassicality, as proposed by Sperling \textit{et al.} in Phys.~Rev.~Lett.~{\bf 109}, 093601 (2012). Our work paves a way for the practical characterisation of increasingly sophisticated states and detectors.
\end{abstract}

\maketitle

States of the electromagnetic field with no analogue in a classical theory of electromagnetism provide a means to characterise quantum coherence and are a valuable resource for quantum-enhanced technologies.  Determining what constitutes a nonclassical state not a trivial matter. Many criteria to establish where the quantum-classical boundary lies have been proposed. Typically, one considers pathological behaviours of distributions in phase space: one looks for negative values of the Wigner quasi-probability function on the quadrature space~\cite{Wig32}, or values more singular than a delta function of the Glauber-Sudarshan $P$ distribution on the phase-amplitude space~\cite{Gla63,Sud63,Vog06}, with the latter definition covering a larger class of states. When detecting photons directly, the Mandel parameter $Q_M$ is often used~\cite{Man79}. This neatly captures any sub-Poissonian behaviour in the photon statistics as a witness of nonclassicality.

Experimentally, these criteria have been widely adopted, and inferred either by full reconstruction of the distribution~\cite{Lvo04,Zav04,Our06,Our07,Zav07}, or by direct measurement~\cite{Lai10,Mar11}. A different approach consists of examining the photon number distribution in a counting experiment~\cite{Man79,Sho83,Rar87,Bon05,URe05,Bus08,Ave08}. While actual photon number resolving detectors are becoming more frequently available~\cite{Wak04,Ger10,Ger11,Bri12}, simpler solutions are still appealing as quantum networks become more complex. Multiplexed detectors are a commonplace choice for accessing higher order Fock states by binning the light to several on-off detectors~\cite{Ach03, Lun08}.  As such, these detectors do not output the photon number distribution. It is recognised that the Mandel parameter applied to the click distribution would give false indications of nonclassicality even for a large number of detection bins~\cite{Spe11}. A linear inversion technique is then required~\cite{Wor09,Lai10,Ave10,Bar12} in order to obtain the actual distribution from the distribution of the clicks. However, this becomes more sensitive to noise and less tractable with increasing system size.

\begin{figure}[b]
\centerline{\includegraphics[width=1 \linewidth]{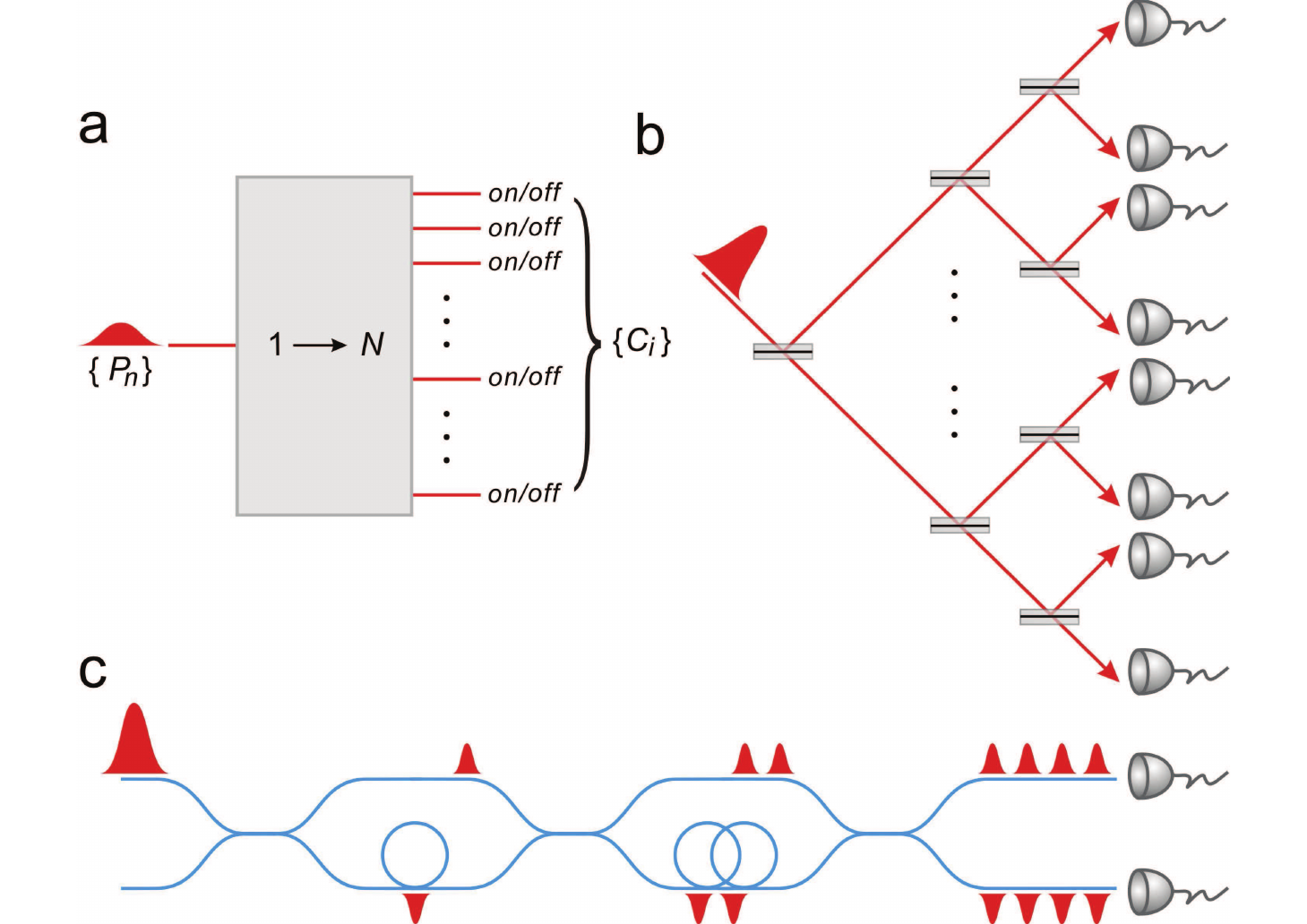}}
\caption{(Color online) a: General scheme of a multiplexed detector. Access to more detail of the photon distribution is obtained by dividing the input beam onto several modes, which are then measured with on/off detectors, such as APDs. The photon number distribution $\{p_n\}$ determines the statistics of the counts $\{c_i\}$, {\it i.e.} the probabilities of observing events with $i$ clicks regardless the bin in which they occur. b: Spatially multiplexed detector: the input beam is split by a set of beam splitters onto a collection of spatial modes. c: Time-multiplexed detector: the input beam is coupled in a single-mode fibre and split on a first 50:50 beam splitter. The two arms are then recombined to form an unbalanced Mach-Zehnder interferometer with a relative delay $\Delta t{=}$\unit[50]{ns}. This produces four distinct modes: two time bins on two distinct spatial modes. This operation is repeated in a second unbalanced interferometer with double relative delay $2\Delta t$, in order to produce two more time bins.  This is the detector we adopted in the present experiments.}
\label{fig:tmd}
\end{figure}
In this paper, we directly observe a signature of nonclassicality in the output of a time-multiplexed detector, following the proposal in Ref.~\cite{Spe12} which identifies a sub-binomial behaviour in the click statistics. We experimentally measure this sub-binomial character for three families of quantum states and compare it with the standard Mandel parameter~\cite{Man79}, as well as a na\"ive analysis of the data. Our results show that the sub-binomial behaviour of click statistics is a reliable indicator of nonclassicality of single-mode light fields. Moreover, this test performs just as well at inferring the true Mandel parameter from the photon distribution. As quantum information systems evolve in sophistication, characterisation via full tomography becomes prohibitively expensive, and our results pave the way for a simple characterisation of quantum states and detectors.

The general scheme of a multiplexed detector is illustrated in Fig.~1a. These simple devices permit access to more detail about the number distribution of a light field through sequential division of an input beam on a series of beam splitters, with detection of each output mode by avalanche photodiodes (APDs). Different architectures have been realised by using either spatial (Fig.~1b)~\cite{Kok01} or time-bin (Fig.~1c)~\cite{Ban03} multiplexing.  In this way, APDs with no photon number resolving capabilities can still deliver information about an input state with $n{>}1$ photons. However, this is non-deterministic, occurring only when the photons split into $N$ separated spatial or temporal bins. This means that the distribution of the photons $\{p_n\}$, and the one of the clicks from the APDs $\{c_i\}$ are different, and their difference only scales with the number of bins as $1/N$~\cite{Spe11}. Some characterisation of the device is required to find the matrix relating the two distributions. Since the splitting operates a linear transformation, the matrix can be inverted in order to obtain the values of $\{p_n\}$~\cite{Lun08,Lai10,Zha12}. In our realisation, we adopt a time-multiplexed detector (TMD) which can detect up to $N{=}8$ events by splitting the input into two spatial and four temporal modes, whose separation is larger than the typical dead time of an avalanche photodiode \cite{Ach03,Lun08,Ave10,Bar12}.

The question arises whether there exists a way of observing nonclassicality directly in the measured quantities $\{c_i\}$. A solution can be found by inspecting the Mandel parameter:
\begin{equation}
\label{Mandel}
Q_M=\frac{\Delta^2n}{\bar n}{-}1.
\end{equation}
The negativity of this parameter establishes a sufficient criterion for nonclassicality: this implies that the mean $\bar n{=}\sum_{n=0}^\infty np_n$ is larger than the variance $\Delta^2 n{=}\sum_{n=0}^\infty {(n-\bar n)}^2p_n$, {\it i.e.} if the statistics are sub-Poissonian. Coherent states, which benchmark classical light, display Poissonian number statistics. By contrast, Fock states display sub-Poissonian statistics, hence the corresponding Mandel parameter takes negative values. When using TMDs, a Poissonian distribution of the photons $\{p_n\}$ results in a binomial distribution of the counts $\{c_i\}$ \cite{Spe11, Spe12}. Based on this, Sperling, Vogel and Agarwal (SVA) have developed a different nonclassicality criterion applying to the measured click statistics:
\begin{equation}
\label{SVA}
Q_B=\frac{\Delta^2c}{\bar{c}(1-\frac{\bar c}{N})}{-}1,
\end{equation}
where $\bar c=\sum_{i=0}^N i\,c_i$, and $\Delta^2 c{=}\sum_{i=0}^N (i-\bar c)^2c_i$. This extends Mandel's formula~\eqref{Mandel} to the case of a binomial distribution. Consequently, nonclassicality can be directly inferred from the sub-binomial behaviour of the observed statistics, {i.e.} $Q_B<0$.

Here, we experimentally measure both $Q_B$ and $Q_M$ for families of quantum states that may be tuned continually between the classical and nonclassical regimes. To obtain such states, we interfere a single photon with a coherent state on a beam splitter with variable reflectivity $R$, and consider the state on the transmitted arm conditioned on the presence of $k$ photons on the reflected arm~\cite{Wat88,Agar89}. For $k{=}1$, we produce non-Gaussian states by performing photon catalysis \cite{Bar12, Lvo02}. These allow us to investigate the behaviour of $Q_B$ in the experiment, as we tune the state from a single photon ($R{=}1$) to a coherent state ($R{=}0$).

\begin{figure}[t!]
\centerline{\includegraphics[width=1 \linewidth]{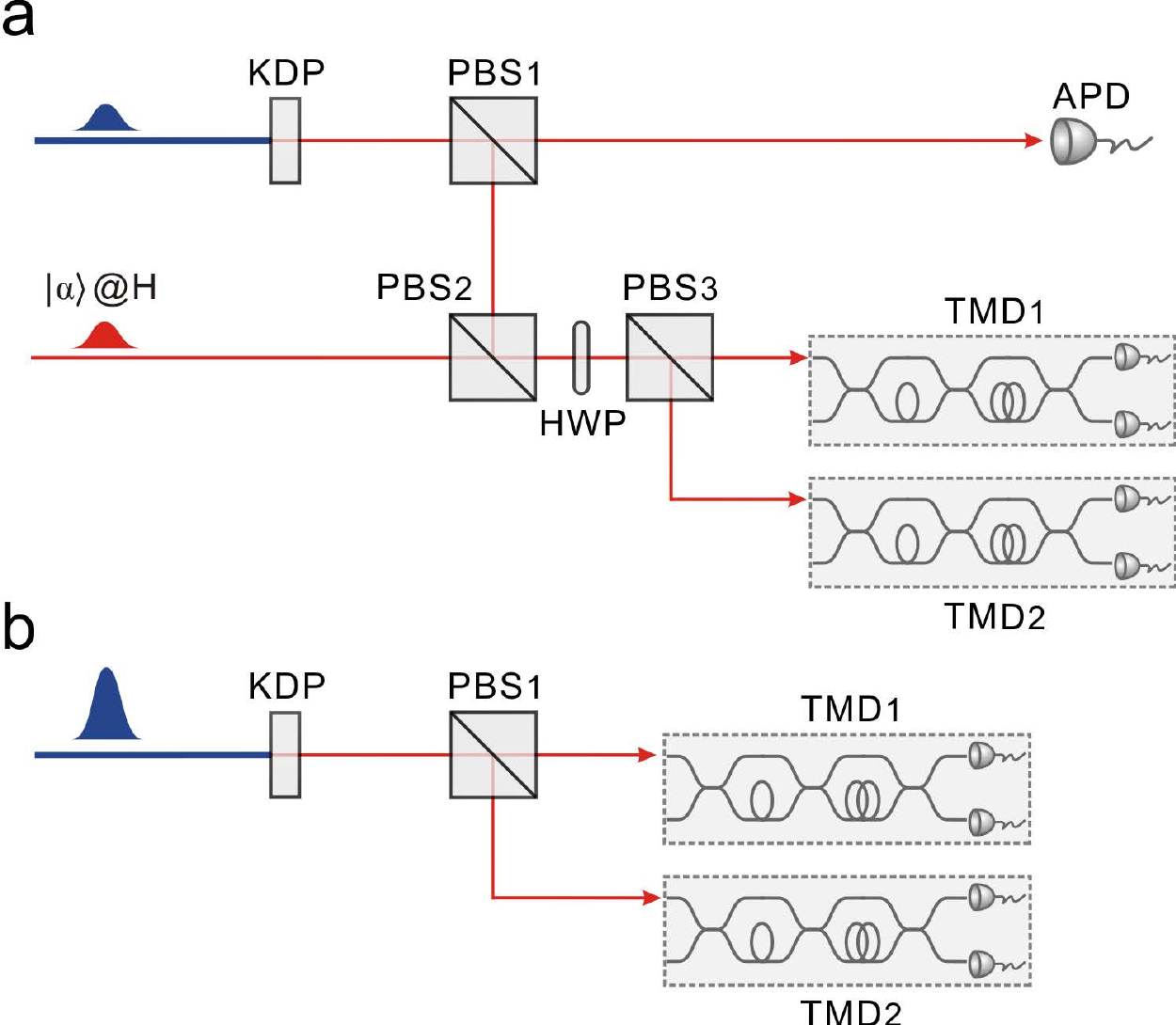}}
\caption{(Color online) a: Measurement of the nonclassicality of a single-mode state. Photons are produced by pulsed type-II co-linear parametric down-conversion in a KDP crystal driven by a doubled Ti:Sa pump laser ($\lambda_{\text p}{=}$\unit[415]{nm}, $\Delta\lambda=$\unit[50]{nm}, repetition rate \unit[256]{kHz}). This produces pairs of in a single spatio-temporal mode~\cite{Mos08}. The presence of a photon in the $V$ mode reflected by PBS$_1$ is heralded by a click of an APD on the $H$ mode, and a trigger derived from the driving laser. A coherent state $\ket{\alpha}$, derived from the main laser, is made to interfere with the single photon on a variable beam splitter consisting of PBS$_2$, a half-wave-plate HWP, and PBS$_3$. The effective reflectivity of the device is then set by the angle $\theta$ of the HWP as $R=\cos^2\theta$. The transmitted arm is delivered to detector TMD$_1$ which provides a second herald event, either no clicks $k{=}0$ or a single click $k{=}1$. The conditional states on the reflected arm are then analysed by TMD$_2$. b: Observing sub-binomial behaviours in the joint click statistics from a two-mode squeezed state generated from the same crystal at higher pump intensity. The two modes from the KDP are directly delivered onto two detectors TMD$_1$ and TMD$_2$, and clicks are collected.}
\label{fig:exp}
\end{figure}

The experimental scheme is shown in Fig. \ref{fig:exp}b: a single photon is produced by pulsed parametric down-conversion in a nonlinear KDP crystal. This interaction produces a pair of photons with orthogonal polarisations in a spectrally-factorable state \cite{Mos08}: the horizontally ($H$) polarised photon is detected by an APD and serves as a herald event. The vertically polarised ($V$) photon is spatially overlapped with a coherent state $\ket{\alpha}$, $H$-polarised, on a polarising beam splitter (PBS). A variable beam splitter, comprising a half-wave-plate and a second PBS, is then used to interfere the two beams. We monitor one output port ,heralding on the measurement of either $k{=}0$, or {$k{=}1$} photons by a first TMD. The conditional state on the second port constitutes our quantum signal, and the complete click statistics from a second TMD $\{c_i\}$ are collected.

\begin{figure}[th]
\centerline{\includegraphics[width=1 \linewidth]{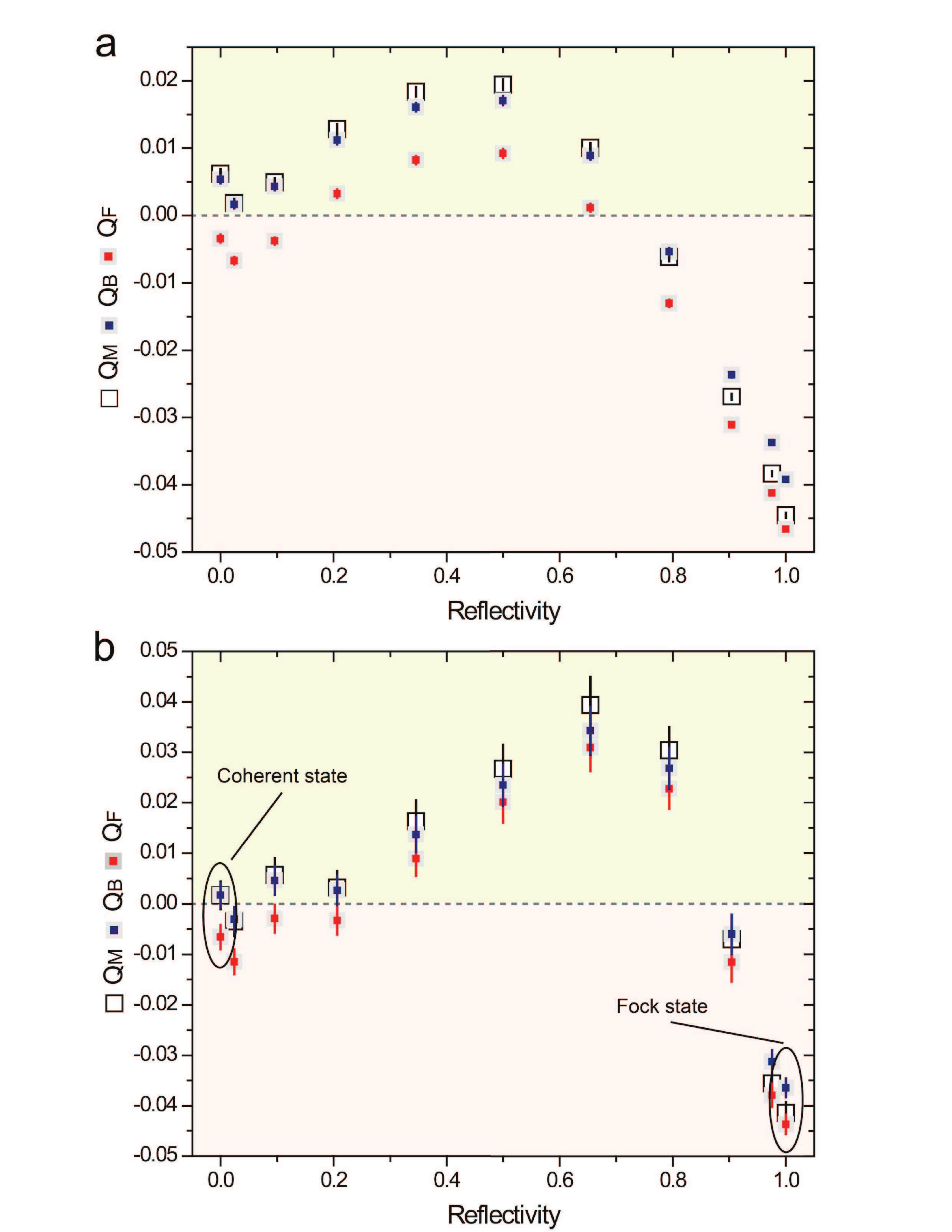}}
\caption{(Color online) Test of nonclassicality for single-mode states heralded by (a) $k{=}0$, and (b) $k{=}1$. In both figures, the black boxes show the the Mandel parameter $Q_M$ and the blue points show the SVA parameter $Q_B$. The values of the Mandel-like parameter calculated from the clicks $Q_F$ are shown as red points. The limit $R\sim0$ correspond to coherent states for which one expects no nonclassicality, while around $R\sim1$, the states approximate displaced single-photons~\cite{Bar12}. All experimental points are obtained by 600 s of data collection. Errors are evaluated by a Monte Carlo technique assuming Poissonian noise.}
\label{fig:k0}
\end{figure}

We perform three different statistical analyses on the collected data, whose results are reported in Fig.~\ref{fig:k0}a for $k{=}0$, and in Fig.~\ref{fig:k0}b for $k{=}1$. In the first analysis, we assess the sub-Poissonian character of the click statistics using a Mandel-like parameter (red squares) defined as
\begin{equation}
Q_F{=}\frac{\Delta^2c}{\bar c}{-}1.
\end{equation}
The blue dots report the experimental result for the genuine Mandel parameter Eq.~\eqref{Mandel}, after the photon distribution has been obtained by inversion~\cite{Wor09}. Finally, the experimental values for the binomial parameter Eq.~\eqref{SVA} are shown as the black squares.

\begin{figure}[t]
\centerline{\includegraphics[width=1 \linewidth]{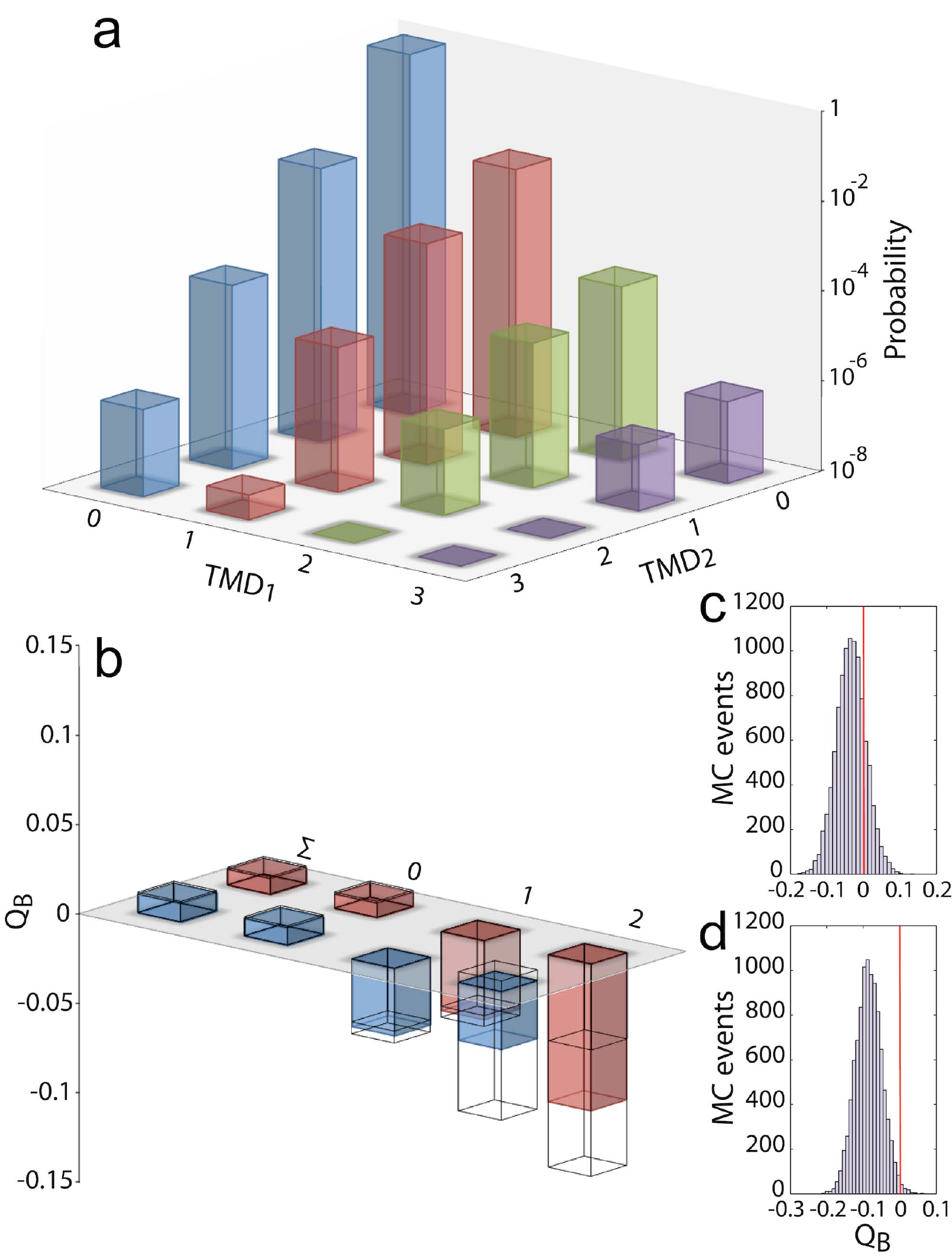}}
\caption{(Color online) a. Joint statistics for detection with TMD on two-mode squeezed states, due to limited efficiency ($\sim7\%$), the most frequent event is when no clicks are registered at either side. All the events in coincidence with a trigger signal from the main laser are included; b. Measured $Q_B$ parameter for the state detected by TMD$_2$ conditioned on different click events on TMD$_1$ c. Statistical analysis of the $Q_B$ parameter when conditioning on two clicks from TMD$_2$ (left), and TMD$_1$ (right). The histograms are obtained based on $10^4$ Monte Carlo simulations assuming Poissonian noise.}
\label{last}
\end{figure}

In general, the Mandel-like parameter $Q_F$ cannot be used to detect nonclassicality, as it can take negative values even for the coherent states obtained when $R{\sim}0$ \cite{Spe11}, as shown in Fig.~\ref{fig:k0}b. Further, its behaviour depends on the particular class of states under investigation: for the case in Fig.~\ref{fig:k0}a the separation between this parameter and the Mandel parameter can be significant, while they are closer in Fig.~\ref{fig:k0}b. In both cases, a systematic failure of $Q_F$ can be seen at low reflectivities. This issue is resolved by employing the binomial parameter $Q_B$, which performs as well as the Mandel parameter $Q_M$ for our eight-bin detector, and with similar uncertainties. In our experiment, we have investigated an intensity regime in which the inversion is not affected by the high-order time bins, in order to have a reliable quantity of comparison for the SVA parameter of our states. However, this might not be feasible in the general case, as uncertainties in these terms will propagate badly in the reconstructed $\{p_n\}$. Thus the direct approach of SVA is useful in avoiding this source of error, as well instabilities arising from noise.

We also use the SVA parameter to characterise the output from the down-converter at higher gain, producing a two-mode squeezed state of the form $\ket{\psi}{\propto}\sum_{n\geq0}\lambda^n\ket{n}_H\ket{n}_V$ on the two orthogonally polarised output modes. If we were to use a photon-number resolving detector, the observation of $n$ photons on the $H$ mode would imply the presence of $n$ photons, hence a sub-Poissonian state, on the $V$ mode \cite{Col09,Dov12}. In our experiment, as shown in Fig.~2b, we observe a similar behaviour in terms of sub-binomial light: a click from a TMD projects the other mode on to a sub-binomial state. The joint statistics from two TMDs, each measuring a different polarisation mode is shown in Fig.~\ref{last}a: from this we have performed the SVA test on the states obtained by conditioning on a given number of clicks on one TMD or the other.

When all the events of TMD$_2$ are considered regardless of the measurement result at TMD$_1$, or {\it vice versa}, we observe classical behaviour, as expected for a thermal state (Fig.~\ref{last}b): we obtain the values $Q_B=(9.3\pm0.6)\cdot10^{-3}$ for the state on mode 1, and $Q_B=(10.9\pm0.6)\cdot10^{-3}$ for the state on mode 2. Discrepancies with respect to the expected symmetry can be attributed to the different detection efficiencies which have been measured. When only the events causing no clicks on TMD$_1$ are registered, we still observe a similar behaviour: the reduced efficiency of our detector, in fact, makes this event similar to tracing out the other mode. The measured values are $Q_B=(10.2\pm0.6)\cdot10^{-3}$ (mode 1) and $Q_B=(8.5\pm0.6)\cdot10^{-3}$ (mode 2). When single clicks are considered as heralds, the state becomes sub-binomial, and so it happens independently of the mode we use as the trigger: $Q_B=(-3.84\pm0.33)\cdot10^{-2}$ (mode 1), and $Q_B=(-4.49\pm0.30)\cdot10^{-2}$ (mode 2). Finally, when triggering on two clicks, the state does present some evidence for nonclassicality, $Q_B=(-3.3\pm3.9)\cdot10^{-2}$ (mode 1) and $Q_B=(-8.3\pm3.5)\cdot10^{-2}$ (mode 2), although the low counts allow us to be confident only for mode 2. A statistical analysis conducted by simulating experiments using a Monte Carlo technique is reported in Fig.~\ref{last}c: while the most likely values for both modes do lie well below zero, the tails of the distribution make the value of $Q_B$ for mode 1 still compatible with positive values.

Our experimental results, in the light of the retrodictive approach in \cite{Amr11}, also provides preliminary evidence for a test of nonclassicality of the TMDs themselves. In the retrodictive framework, the nonclassicality of a detector $D$ can be ascertained by observing the nonclassicality of one mode of $\ket{\psi}$ when the other mode is projected by the action of $D$ in the limit of high squeezing $\lambda{\rightarrow}1$. Although we are far from this regime, the observed negativity of $Q_B$ resulting from conditioning one of the arms on a TMD points towards the nonclassicality of the device.

The direct experimental observation of the nonclassicality of light fields is a vital aspect of foundational studies and technological applications of quantum mechanics. In this work we have shown how this is made possible by the use of time-multiplexed detectors, and by directly inspecting the resulting click statistics. This represents a more elegant and unambiguous procedure for confirming quantumness than their indirect verification by deconvolving the action of the detector.

We thank Brian Smith, Josh Nunn, Steve Kolthammer, and Werner Vogel for discussion and comments. We are also grateful to Arturo Lezama for identifying an error in Eq.~\ref{SVA}, which has since been corrected in this version of the manuscript. This work has been supported by the Royal Society, EU IP Q-ESSENCE (248095), EPSRC (EP/J000051/1 and EP/H03031X/1), AFOSR EOARD (FA8655-09-1-3020). XMJ is supported by an EU Marie-Curie Fellowship (PIIF-GA-2011-300820).

\end{document}